\PassOptionsToPackage{dvipsnames}{xcolor} 
\documentclass[%
 reprint, 
 amsmath,amssymb,
 aps, physrev,
]{revtex4-2}

\usepackage{graphicx}
\usepackage{dcolumn}
\usepackage{bm}
\usepackage{xcolor}
\usepackage{float}
\usepackage{parskip}
\usepackage[dvipsnames]{xcolor}
\usepackage[normalem]{ulem}
\usepackage{soul}
\usepackage[utf8]{inputenc} 
\usepackage[T1]{fontenc} 
\usepackage{hyperref}

\begin{document}

\preprint{APS/123-QED}

\title{\textbf{Learning disentangled latent representations facilitates discovery and design of functional materials} 
}%

\author{Jaehoon Cha}
\thanks{These authors contributed equally to this work.}
\affiliation{Scientific Computing Department, Rutherford Appleton Laboratory, Science and Technology Facilities Council, Harwell Campus, Didcot, OX11 0QX, UK}

\author{Tingyao Lu}
\thanks{These authors contributed equally to this work.}
\affiliation{Department of Chemistry, Physical and Theoretical Chemistry Laboratory, University of Oxford, Oxford, OX1 3QZ, United Kingdom}

\author{Matthew Walker}
\affiliation{Department of Chemistry, University College London, WC1H 0AJ, United Kingdom.}

\author{Keith T. Butler}
\affiliation{Department of Chemistry, University College London, WC1H 0AJ, United Kingdom.}
\email{k.t.butler@ucl.ac.uk}  

\date{\today}

\begin{abstract}
The discovery of new materials is often constrained by the need for large labelled datasets or expensive simulations. In this study, we explore the use of Disentangling Autoencoders (DAEs) to learn compact and interpretable representations of spectral data in an entirely unsupervised manner. We demonstrate that the DAE captures physically meaningful features in optical absorption spectra, relevant to photovoltaic (PV) performance, including a latent dimension strongly correlated with the Spectroscopic Limited Maximum Efficiency (SLME)—despite being trained without access to SLME labels. This feature corresponds to a well-known spectral signature: the transition from direct to indirect optical band gaps. Compared to Principal Component Analysis (PCA) and a $\beta$-Variational Autoencoder ($\beta$-VAE), the DAE achieves superior reconstruction fidelity, improved correlation with efficiency metrics, and more compact encoding of relevant features. We further show that the DAE latent space enables more efficient discovery of high-performing PV materials, identifying top candidates using fewer evaluations than both VAE-guided and random search. These results highlight the potential of DAEs as a powerful tool for unsupervised structure–property learning and suggest broad applicability to other areas of materials discovery where labeled data is limited but rich structure is present in raw signals.
\end{abstract}

\maketitle


\section{\label{sec:intro}Introduction}

Discovering and designing new materials is fundamental to the development of human society.\cite{cheetham2022chemical} New materials are particularly important in the context of climate change and energy transition where more stable, abundant and efficient materials are required for almost every application. The discovery of new materials has often been a slow process, the transition from bronze to iron took in the region of 1500 years! Scientific progress and understanding has resulted in a dramatic increase in the rate at which new materials for a desired application can be developed; however many feel that the rate is too slow to effectively combat the pressing issues of climate change. The advent of materials databases and the rise of machine learning (ML) has raised the prospect of a new data-led acceleration in the rate of materials discovery.\cite{hey2009fourth, butler2018machine} One of the challenges in the era of high-throughput computation and large materials databases is how best to navigate the space of new materials to find promising candidates and how to unlock new design principles from this data. 

The discovery of new materials for photovoltaics (PV) is a prime example of a modern materials discovery challenge. As global PV capacity surpassed 1.6 TW in 2023, \cite{IEAPVPS2024} the projected expansion to 30–70 TW by 2050 demands the identification of novel, efficient absorber materials. \cite{haegel2019terawatt} Traditional approaches—based on experimental synthesis and first-principles calculations—remain essential, but the increasing availability of large materials databases now presents new opportunities to accelerate discovery using machine learning (ML)\cite{scheidgen2023nomad, kirklin2015open, jain2013commentary, choudhary2020joint}.

A growing ecosystem of high-throughput workflows and data infrastructures\cite{ong2013python, larsen2017atomic, pizzi2016aiida} has enabled the generation and aggregation of vast materials datasets. The challenge now lies in effectively navigating these high-dimensional datasets to identify candidates with desirable properties. Here, ML offers powerful tools for pattern recognition, prediction, and optimization that can significantly reduce the time and cost of screening.

ML has already demonstrated its utility across a range of tasks in materials science, including property prediction \cite{chibani2020machine, devi2024optimal, ward2016general, antunes2023predicting}, structure generation \cite{kim2020generative, antunes2023predicting, jiao2023crystal}, inverse design \cite{park2024, noh2020machine}, and interpreting experimental data.\cite{wong2024predicting, mirza2024elucidating} However, a central challenge remains: learning robust and generalizable representations of complex materials data. \cite{rudin2022interpretable, oviedo2022interpretable} High-quality representations are critical not only for predictive performance, but also for interpretability, data efficiency, and transferability to new tasks .\cite{bengio2013representation, spurek2020non}

One particularly important property of learned representations is disentanglement—where individual latent features correspond to independent generative factors of variation. Disentangled representations have shown promise in fields such as scene understanding and generative modelling, \cite{eslami2018neural} as they provide more interpretable and modular representations of underlying data structure. In the context of materials discovery, disentanglement holds the potential to bridge the gap between data-driven models and human-understandable insights, paving the way toward interpretable ML approaches that not only predict properties but also help explain why certain materials perform well.

Autoencoder (AE)-based models have evolved into specialised architectures to address various challenges in representation learning.\cite{hi06:_ae, ki13:_vae,  cha23:_dae} Once AE-based models are trained, their encoder can be used independently to map input data to a low-dimensional latent space. This latent representation can be leveraged for downstream tasks such as clustering or data compression for further analysis.\cite{br23:_versatile, ba23:_aecomp} The decoder can also be used separately to explore how variations in the latent representation affect the original data space, providing insights into the relationship between latent factors and properties of the input dataset. This enables a deeper understanding of the underlying factors driving variations in the dataset.

In this work, we apply the recently proposed Disentangling Autoencoder (DAE)~\cite{cha23:_dae} to explore a large database of candidate photovoltaic materials. The DAE offers a theoretically grounded approach to learning independent, multidimensional subspaces, yielding disentangled representations of material properties. We show that this framework can uncover key aspects of optical absorption spectra in an unsupervised manner, enabling efficient navigation of high-dimensional materials datasets to identify promising photovoltaic absorbers. To demonstrate this, we simulate a discovery campaign using the DAE’s latent space to propose new PV materials, and find that it significantly outperforms random sampling. In addition, we show that DAE-derived representations improve reconstruction accuracy for spectral properties in unseen materials. These findings highlight the potential of disentangled latent representations for accelerating materials discovery, and position DAEs as a powerful tool for data-driven exploration in chemistry and materials science.

\section{\label{sed:data}Dataset}

The dataset we are working with consists of simulated optical absorption spectra of 17283 materials. The spectra are compiled from two datasets of calculated optical spectra. \cite{data-2-1, data-2-2} We calculate a maximum photovoltaic efficiency as described in detail elsewhereusing the spectroscopically limited maximum efficiency (SLME) approach.\cite{yu2012identification} The SLME combines an optical absorption spectrum with the solar spectrum at the earth's surface and using arguments of detailed balance, derives the maximum power conversion efficiency (PCE) obtainable by a single junction solar cell based on that material as an absorber layer. Maximising SLME is often used as a target when searching for promising new PV materials.\cite{walker2025carbon}

\begin{figure*}[ht!]
    \centering
    \includegraphics[width=0.75\textwidth]{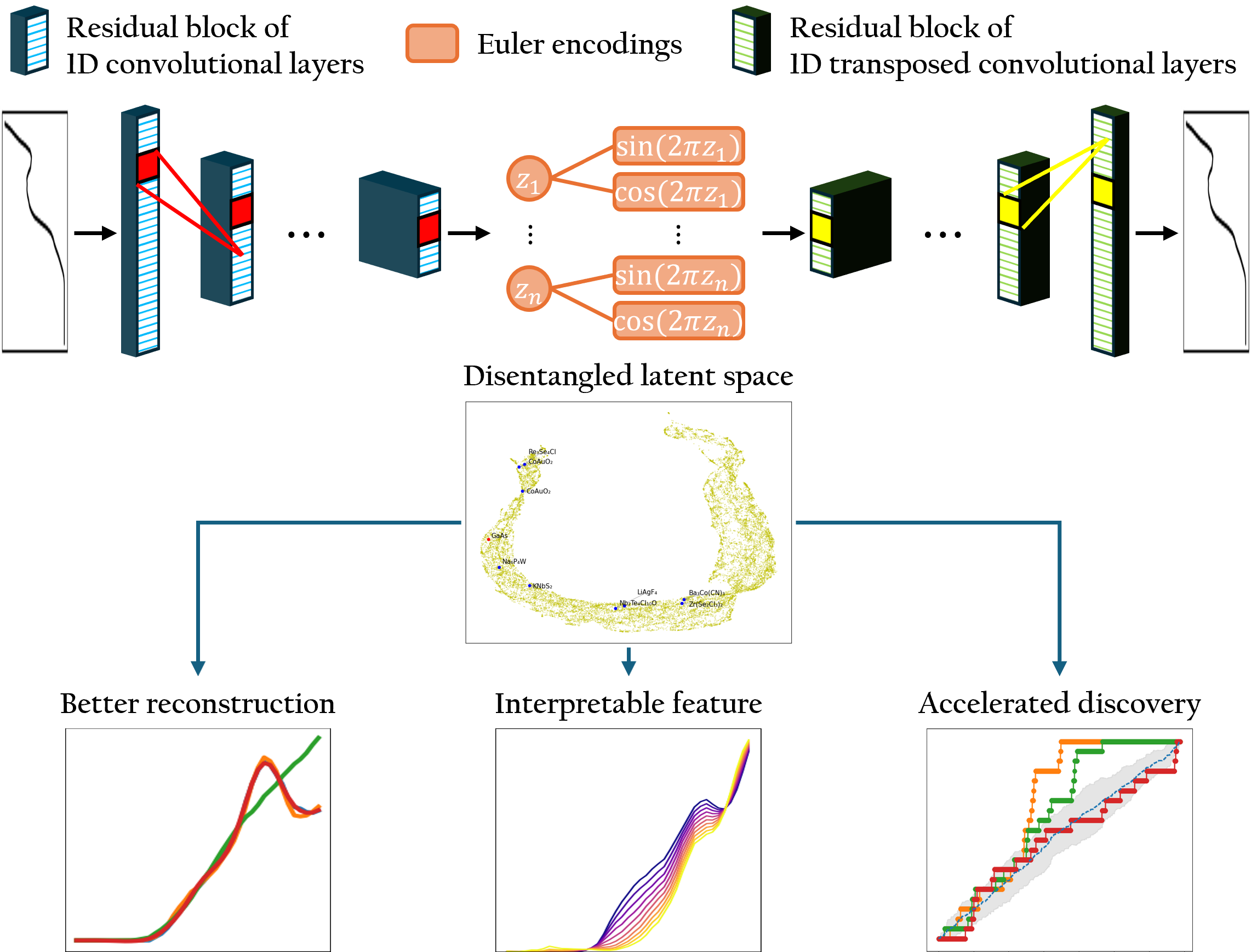} 
    \caption{A depiction of the disentangled autoencoder used in this study. Residual layers and Euler layers are used to encode a disentangled latent space, which we show to provide better quality, more interpretable representations of the optical absorption spectra.}
    \label{fig:overview}
\end{figure*}

\section{\label{sec:res}Results and discussion}

The Disentangling Autoencoder (DAE), shown in Figure~\ref{fig:overview}, offers a novel approach to learning latent representations by enforcing orthogonality in the latent space, ensuring that each latent dimension corresponds to a distinct and interpretable source of variation. \cite{cha23:_dae} In contrast to probabilistic models such as VAEs.\cite{ki13:_vae} the DAE explicitly disentangles these variations by enforcing orthogonality through its architectural design (see Figure~\ref{fig:overview}), enabling more direct and interpretable representations while avoiding common issues like posterior collapse. While DAEs have shown strong performance in extracting features from 2D image data,\cite{cha23:_dae, cha25discovering} here we adapt the approach to 1D signals, designing a 1D convolutional DAE tailored to capture key features in the optical absorption spectra of materials.

In our workflow, we train a DAE with a 9-dimensional latent space using only a reconstruction loss, aiming to capture the most salient and independent sources of variation within the dataset of optical absorption spectra. The encoder comprises residual blocks, each containing two 1D convolutional layers, followed by fully connected layers. The decoder mirrors this structure, with fully connected layers followed by residual blocks. Unlike VAEs, the DAE does not rely on a Kullback–Leibler divergence term. Instead, it promotes disentanglement through a combination of normalisation, interpolation, and the Euler layer,\cite{cha23:_dae} which constrains the decoder’s output variations to be orthogonal across latent dimensions. Full details of the model architecture are provided in the methods section, and the models are available online.

To evaluate the quality of the learned representations, we compare the DAE with a $\beta$-Variational Autoencoder ($\beta$-VAE), which is a widely used baseline for  learning disentangled features, and principal component analysis (PCA). Both DAE and $\beta$-VAE are trained on the same optical absorption dataset, with identical encoder and decoder architectures and the same nine-dimensional latent space. In the $\beta$-VAE, the $\beta$ coefficient encourages disentanglement in the latent space, \cite{bu18:_bvae} offering a conceptually comparable baseline to the DAE, which achieves disentanglement through architectural design. For PCA, we reduced the data to nine principal components that matches the dimensionality of the latent space used in DAE and $\beta$-VAE. We applied the fit transform and inverse transform functions to project the optical data into nine-dimensional vectors and subsequently reconstruct the original data from these representations.

We assess model performance along two complementary axes: reconstruction fidelity and the quality of learned features. The former evaluates whether the latent representation retains sufficient information to accurately describe the original spectra, while the latter probes whether the latent dimensions capture physically meaningful variations relevant to photovoltaic efficiency. Together, these criteria indicate a model’s ability to learn important features while maintaining good reconstruction. We start by assessing the meaningfulness of the learned representations by simulating a materials discovery campaign, and then by the latent traversal method. We then compare the reconstructions from DAE, VAE and PCA to compare the fidelity of the learned features in terms of how much information they preserve from the original spectrum.

\subsection{DAE facilitates materials discovery}

\begin{figure}[ht!]
    \centering
    \includegraphics[width=0.49\textwidth]{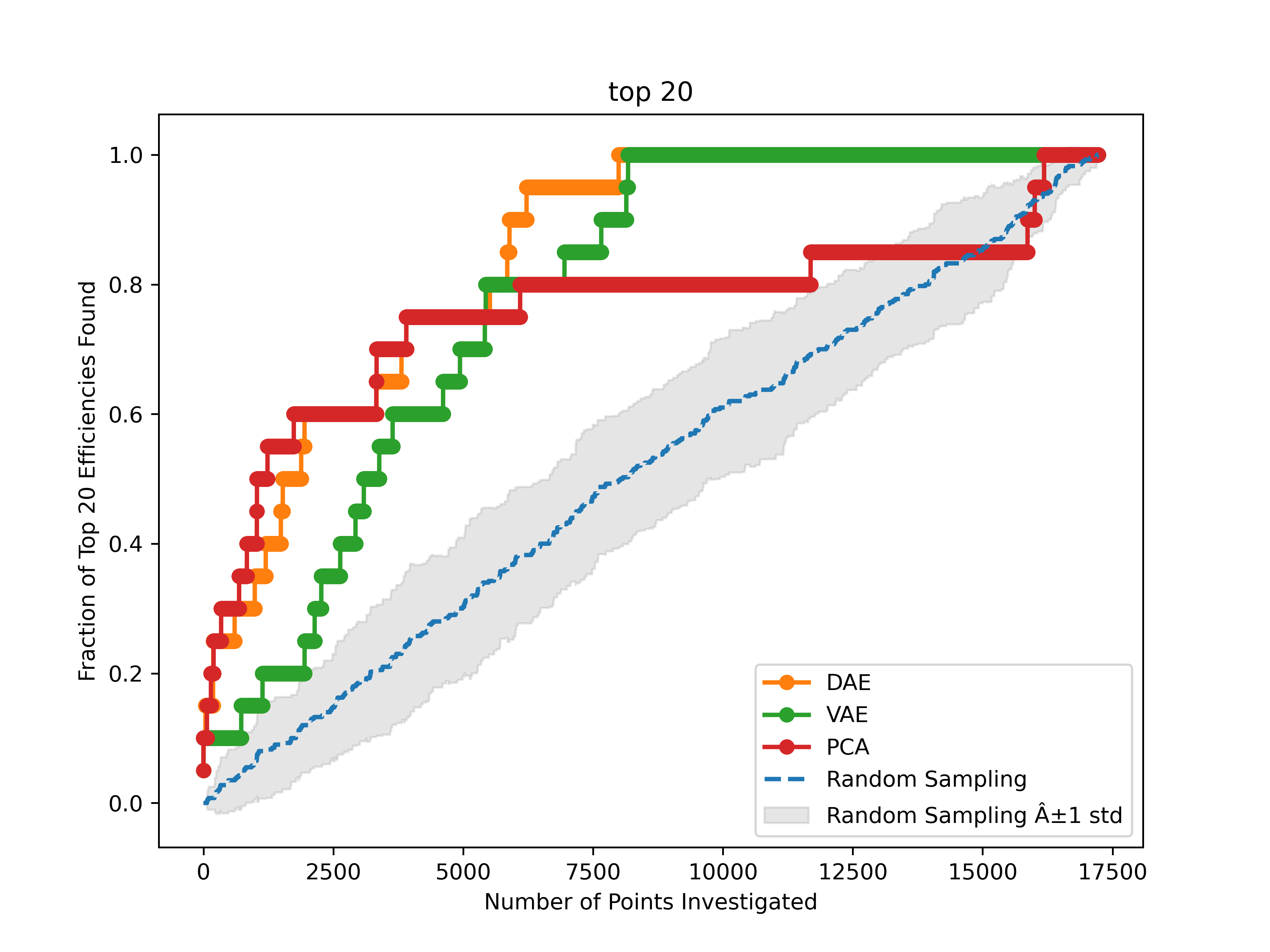} 
    \caption{Discovery rates for 9 latent dimensions using DAE (orange), VAE (green) and random (red) method with $\pm$ 1 standard deviation (grey). The ordinate axis shows the number of materials visited in the latent space search and the abscissa shows the percentage of the top 20 SLME materials recovered within that set.}
    \label{fig:efficiency-research}
\end{figure}


To demonstrate the practical value of the DAE representations for materials discovery, we simulate a realistic screening scenario. Given the absorption spectrum of a known high-performing photovoltaic material we ask whether the latent representations learned by the DAE can help identify other promising PV candidates in a large, unlabeled dataset. Importantly, this search is conducted without access to PV efficiency labels, which means the model must rely solely on structural patterns in the absorption spectra.

Our approach uses the DAE and $\beta$-VAE latent representations and PCA components to rank candidate materials by the distance from a chosen material with known high SLME in the latent space. The hypothesis is that if the latent space is meaningfully structured, materials close to this known material should exhibit similar photovoltaic potential. We then evaluate how many candidates must be examined in order to discover the top 20 materials ranked by SLME.

The results, shown in Figure~\ref{fig:efficiency-research} highlight the effectiveness of latent space exploration as an approach to materials discovery. All three methods show significant improvements compared to the random sampling baseline. DAE and PCA have recovered more than 60\% of the top 20 materials by exploring less than 15\% of the search space. The $\beta$-VAE exploration is also significantly better than random, but not as effective as the other two methods. After discovering around 75\% of the top 20 materials, the performance of DAE notably exceeds that of the PCA. We conjecture that the ability of DAE to uncover subtle features in the underlying data facilitates this discovery.  DAE has discovered all of the top 20 materials by exploring around 43 \% (7500 out of 17282) candidate materials.

These findings reinforce the broader value of learned and disentangled representations: by isolating interpretable, physically meaningful features, the DAE enables more targeted navigation of high-dimensional datasets. In practical terms, this translates into fewer costly simulations or experiments when searching for new functional materials. As such, DAEs represent a powerful tool for accelerating discovery in photovoltaics and, more broadly, in materials science domains where labeled data is scarce but structured signals exist. As we now demonstrate, the learned reconstructions from DAE not only facilitate discovery, but interpretable features mean that they yield insights into the important features determining performance.

\subsection{Latent traversal reveals important features}

\begin{figure*}[ht!]
    \centering
    \includegraphics[width=\textwidth]{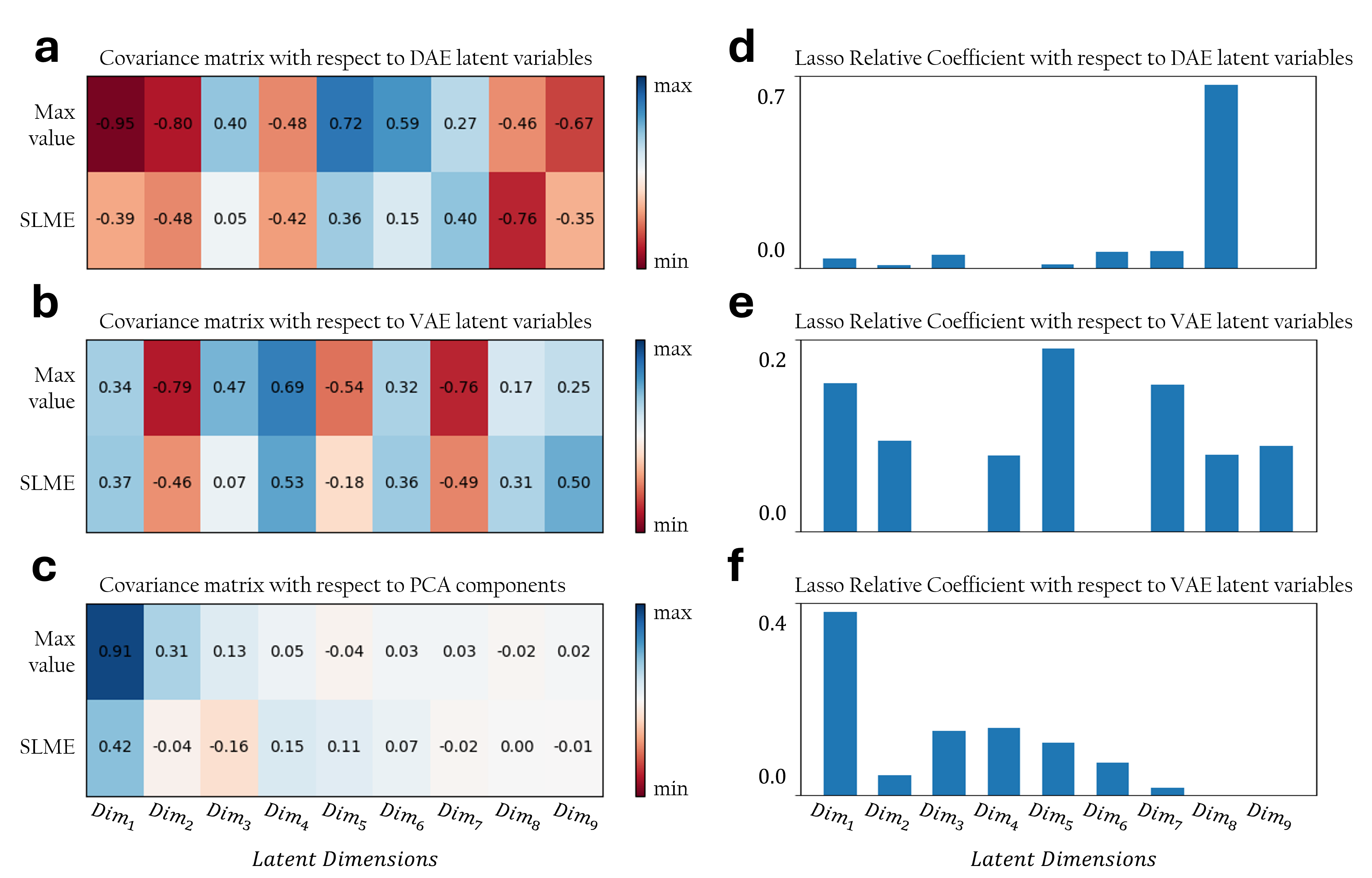} 
    \caption{Covariance matrix of maximum values in the spectrum and the SLME with respect to latent variables obtained using (a-c) DAE, VAE and  PCA. (d-f) The coeffieicnts of the contriution of different latent variables to the Lasso models traind to predict the SLME based on the values of the latent variables.}
    \label{fig:dae_vae_comp}
\end{figure*}

\begin{figure*}[ht!]
    \centering
\includegraphics[width=\textwidth]{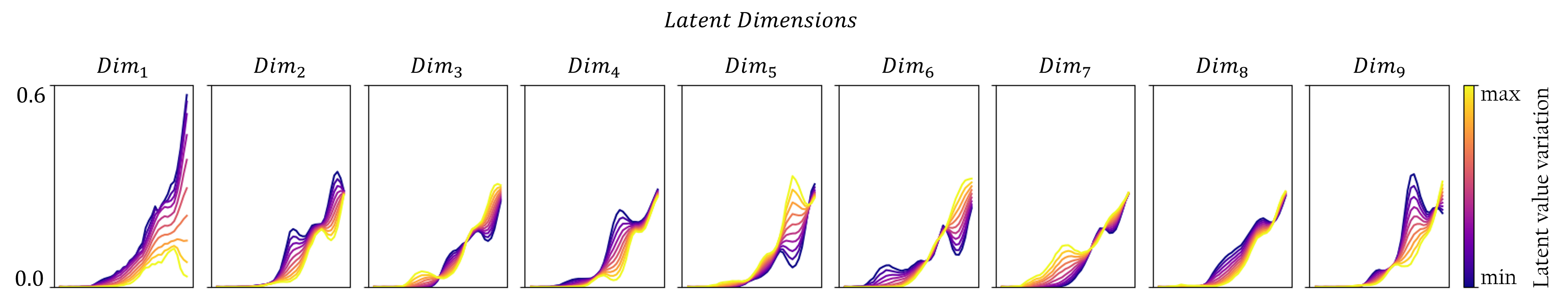} 
    \caption{Latent traversal of all latent dimensions using DAE. For each dimension the latent space is traversed from low latent value to high latent value, keeping all other latent dimensions fixed. The resulting spectra are plotted here and colour coded based on latent dimension value.}
    \label{fig:lat_traversal}
\end{figure*}

\begin{figure}[ht!]
    \centering
    \includegraphics[width=0.45\textwidth]{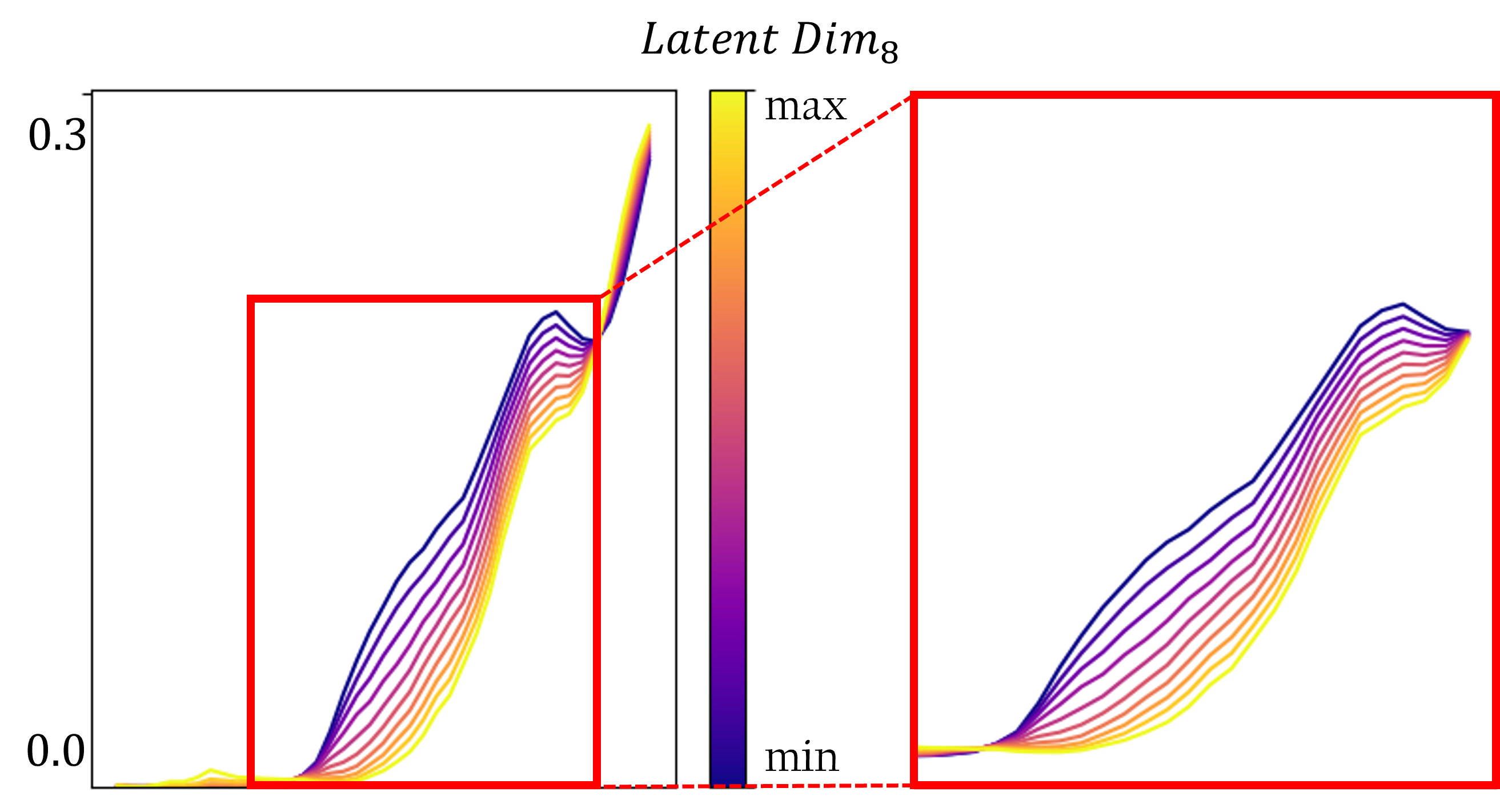} 
    \caption{The zoomed-in view illustrates variations concerning SLME. This dimension exhibits a negative correlation with SLME, therefore the traversal from left to right corresponds to a decrease in SLME. This feature corresponds to the change from a direct band gap (left) to an indirect band gap (right).}
    \label{fig:lat_traversal_slme}
\end{figure}

We now demonstrate that as well as facilitating discovery, disentangled representations yield practical insights for materials design. We show that DAE, trained without any labelled data, is able to identify features in the optical absorption spectrum that are strongly associated with photovoltaic efficiency, as measured by the SLME. The DAE was trained solely on the optical absorption dataset, with no SLME labels provided during training. After training, we computed the Pearson correlation coefficient between each of the nine latent dimensions and SLME (Figure~\ref{fig:dae_vae_comp}a). This analysis revealed one latent dimension with a strong negative correlation of -0.76 with SLME, indicating a meaningful association between the encoded spectral feature and PV performance. For context, a correlation of -1 indicates a perfect negative correlation, while +1 indicates a perfect positive correlation.

Importantly, the SLME-associated latent feature stands out clearly as the only one among the nine latent features with a strong correlation to SLME in the DAE features (Figure~\ref{fig:dae_vae_comp}a ). In addition, it remains largely independent of the other latent features: its absolute correlation with other metrics, such as the maximum value of absorption, remained below 0.5 (Figure~\ref{fig:dae_vae_comp}a), suggesting that the model has successfully isolated an interpretable and disentangled factor linked to PV efficiency.

We performed latent traversals, a technique for visualising what the model has learned, across all latent dimensions to examine how each feature influences the reconstructed spectra. For each traversal, we varied one latent dimension from its minimum to maximum value while keeping all others fixed. The resulting spectra are shown in Figure~\ref{fig:lat_traversal}, where yellow indicates spectra generated using the minimum value of the selected latent feature, and blue corresponds to the maximum. 

To further explore the effect of the SLME-correlated latent feature, we traverse this dimension of latent space shown in Figure~\ref{fig:lat_traversal_slme}. The evolution of the spectrum across this feature dimension exhibits a striking transformation in the shape of the absorption onset: at low values of the latent variable (corresponding to high SLME), the onset is concave, whereas at high values (low SLME), it becomes convex. Physically, this trend is characteristic of a transition from a direct to an indirect optical band gap. Direct gap materials are well known to be more efficient light absorbers than indirect gap materials, which require a phonon-assisted process for absorption. This distinction is explicitly captured in the phenomenological model underlying SLME, yet it is remarkable that the DAE is able to recover this complex, physically meaningful pattern entirely from the absorption spectra and in an unsupervised manner. 
\subsection{DAE learns more relevant information than $\beta$-VAE and PCA}

\begin{figure*}[ht!]
    \centering
    \includegraphics[width=\textwidth]{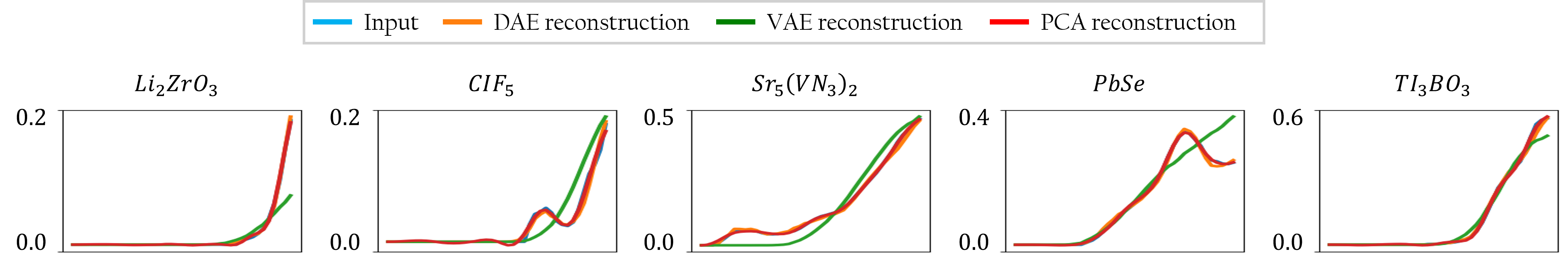} 
    \caption{Reconstruction comparison between DAE, VAE and PCA. The plots show the ground truth spectrum in blue and the spectra reconstructed from the three different dimensionality reduction techniques for five randomly selected systems in the dataset. Note that the absorption values have been normalised. }
    \label{fig:reconstr}
\end{figure*}

In terms of reconstruction accuracy, the DAE, $\beta$-VAE, and PCA achieve mean squared errors (MSE) of 0.00010, 0.00125, and 0.00004, respectively. Although PCA achieves the lowest reconstruction error due to its explicit identification of the linear subspace that minimizes MSE, Figure~\ref{fig:reconstr}, which presents reconstructions of randomly selected optical absorption spectra from all models, indicates that the difference between PCA and DAE reconstructions is negligible. Despite learning a more complex and nonlinear latent representation than PCA, the DAE preserves overall spectral fidelity and captures fine-grained features more effectively than the VAE, as seen in the distinct bump observed in the spectra of materials like PbSe (mp-2201) and ClF$_5$ (mp-1213763). This improvement is notable given that the DAE uses only a reconstruction loss, relying instead on architectural design (i.e., orthogonality constraints) to promote feature disentanglement.

Beyond reconstruction, we evaluate how well each model captures features linked to photovoltaic efficiency. As shown in the previous section, the DAE learns a latent dimension with a strong and interpretable correlation to the SLME metric. In contrast, the VAE and PCA do not yield any single latent dimension with a high correlation to SLME (Figure~\ref{fig:dae_vae_comp}b, c), suggesting a more entangled representation.

To quantify how useful the latent features are for predicting SLME, we train a Lasso regression model on the latent variables. Lasso is a linear model that includes built-in feature selection: it penalises complexity by shrinking less important coefficients to zero, helping us identify which latent dimensions are most relevant for prediction. This provides a simple and interpretable test of whether meaningful structure has been encoded. We train the model on 70\% of the data and evaluate performance on the remaining 30\%, using test set MSE as the evaluation metric. The DAE outperforms both the $\beta$-VAE and PCA in this task, achieving a lower prediction error (MSE of 48.6 for DAE, compared to 68.0 for VAE and 83.6 for PCA), indicating that its latent features carry more relevant information for SLME.

Figure~\ref{fig:dae_vae_comp}d, e, f shows the learned Lasso coefficients, revealing that the DAE’s SLME prediction relies almost entirely on a single latent dimension—consistent with the Pearson correlation analysis. By contrast, the $\beta$-VAE and PCA distribute predictive power across multiple entangled features. This compact and interpretable encoding by the DAE highlights its advantage in learning physically meaningful structure from unlabeled spectra.


\section{\label{sec:conc}Conclusion}

In this work, we have demonstrated the utility of Disentangling Autoencoders (DAEs) for learning meaningful, compact representations of optical absorption spectra, with a focus on accelerating the discovery of photovoltaic (PV) materials. Unlike conventional approaches such as $\beta$-Variational Autoencoders ($\beta$-VAEs), the DAE introduces architectural inductive biases to explicitly disentangle latent factors of variation, leading to more interpretable and physically relevant representations.

We showed that a DAE trained solely on absorption spectra, without access to any labels, is able to identify latent dimensions strongly correlated with photovoltaic efficiency as measured by SLME. These features correspond to meaningful physical phenomena—most notably, the transition from direct to indirect optical band gaps—which are central to the design of efficient PV absorbers. Compared to the PCA and $\beta$-VAE, the DAE not only achieved lower reconstruction error but also encoded SLME more compactly and predictively in its latent space.

Building on this representation, we simulated a discovery campaign starting from a known material, and found that the DAE latent space enabled more efficient identification of top-performing PV candidates than either the PCA, $\beta$-VAE or random search. This highlights the practical value of DAEs in navigating large, unlabelled materials datasets and prioritising candidates for further investigation.

Looking ahead, the potential applications of DAEs extend far beyond photovoltaics. Any domain in which complex, high-dimensional materials data is available—such as X-ray diffraction patterns, density of states, or even molecular dynamics trajectories—could benefit from unsupervised, disentangled representations that capture the underlying physics. In particular, DAE-based latent spaces could serve as a foundation for inverse design, generative modelling, or transfer learning across related materials classes. Overall, our findings suggest that DAEs are a promising tool for structure-property discovery in materials science, offering a pathway to more interpretable and efficient machine learning workflows that operate even in the absence of labels.

\section{\label{sec:meth}Methods}

\textbf{Source of data:} The optical electronic spectra for photovoltaic materials are obtained from the \texttt{ForbiddenTransitions} dataset on the MPContribs platform, which contained 16,533 (at the time of access)  spectra from materials with a diverse range of structures and chemistries,\cite{data-2-1, huck2015community, material-project-2, material-project-3} this was combined with the \texttt{Screening Inorganic PV} dataset also on the MPContribs platform \cite{data-2-2}


\textbf{Discovery method:} To evaluate the practical utility of the learned representations, we simulated a materials discovery campaign using a nearest-neighbour search in latent space. Specifically, we assessed how efficiently the model could identify the top 20 materials by SLME.

 Each material was represented by its latent value, and similarity to the top SLME material in the dataset was computed using Manhattan distance. 
The discovery procedure proceeds by ranking all materials according to their distance from the top SLME material. A nearest-neighbour search is then performed by iteratively evaluating candidates in order of increasing distance. At each iteration, we track the cumulative number of top-performing materials (by SLME) discovered, allowing us to quantify search efficiency. To provide a baseline, we compare this targeted search to a random sampling strategy. Random selections were repeated over multiple trials to compute mean and standard deviation for statistical comparison.

\textbf{Model architecture:} Both the DAE and $\beta$-VAE share the same encoder–decoder architecture. The encoder takes a one-dimensional optical absorption spectrum of size 1×41 as input. It consists of four residual blocks with 8, 16, 32, and 64 channels, respectively. Each block comprises two branches: a main branch and a residual (shortcut) branch.
The main branch begins with a 1D convolutional layer with a kernel size of 3, stride of 2, and padding of 1, followed by 1D batch normalization (BN) and a Leaky Rectified Linear Unit (Leaky ReLU) activation. This is followed by a second 1D convolutional layer with a kernel size of 3, stride of 1, and the same padding, again followed by 1D BN and Leaky ReLU. In parallel, the residual branch applies a single 1D convolutional layer with a kernel size of 3, stride of 2, and padding of 1 directly to the input, followed by 1D BN, to match the dimensionality of the main branch output. The outputs from both branches are then added to form the final output of the block.

In the DAE, a single fully connected (FC) layer follows the last residual block, whereas the $\beta$-VAE includes two FC layers to output the mean and log-variance for reparameterization. Both models use a latent space of nine dimensions. This choice was determined empirically by testing different latent dimensions and selecting the smallest dimension that still allowed the models to accurately reconstruct the original spectra. When the latent dimension was reduced below nine, the models tended to produce overly smoothed or flat spectra, indicating a loss of essential information.

The decoder is constructed symmetrically to the encoder: it begins with a fully connected layer and includes four residual blocks using 1D transposed convolutional layers with 32, 16, 8, and 1 channels, respectively, each with a stride of 2. Each of these blocks follows the same structure as those in the encoder, with both a main and a residual branch. Following the residual blocks, a final 1D convolutional layer with a kernel size of 3 and a Sigmoid activation function is applied to reconstruct the input optical absorption spectrum.

We used the Adam optimizer with a learning rate of 0.0001 and a batch size of 64 for training both models. Each model was trained for 1000 epochs using Mean Absolute Error (MAE) as the loss function. Both DAE and $\beta$-VAE incorporate a hyperparameter to encourage disentanglement, as introduced in their respective formulations: $\alpha$ for the DAE and $\beta$ for the $\beta$-VAE. In our experiments, we set $\alpha$ = 0.005 and $\beta$ = 2.

The models were trained using a single NVIDIA Tesla V100 GPU on the ADA system at the Science and Technology Facilities Council (STFC). The implementation was done in PyTorch, which enabled efficient GPU acceleration. The results are fully reproducible\footnote{Code available at: \url{https://github.com/Jaehoon-Cha-Data/material_discovery/tree/main}}.

\section{Code and data availability}

All code and associated data required to reproduce this work are openly available at \url{https://github.com/619
Jaehoon-Cha-Data/material_discovery/tree/main}.



\section{Acknowledgements}
JC acknowledges support from the Ada Lovelace Centre. KTB acknowledges support from EPSRC project EP/Y000552/1 and  EP/Y014405/1. Via our membership of the UK's HEC Materials Chemistry Consortium, which is funded by EPSRC (EP/X035859/1), this work used the ARCHER2 UK National Supercomputing Service (http://www.archer2.ac.uk).

\bibliography{apssamp}

\end{document}